# Radar Cross Section Measurement within Reverberation Chamber: Stirrer position issues

Ariston Reis[1], François Sarrazin[1], Philippe Pouliguen[2], Jérôme Sol[3], Philippe Besnier[3], Elodie Richalot[1]
[1] Université Paris-Est, ESYCOM (FRE 2028), CNAM, CNRS, ESIEE Paris, F-77454 Marne-la-Vallée, France.
[2] Agence Innovation Défense, Direction Générale de l'Armement, 75509 Paris, France.
[3] INSA Rennes, CNRS, IETR-UMR 6164, F-35000 Rennes.
* ariston.defreitastavaresdosreis@u-pem.fr

*Abstract*—**This paper presents the evaluation of the Radar Cross Section (RCS) of a metallic object by measurements accomplished within the diffuse-field environment produced by a Reverberation Chamber (RC). The method is based on the extraction of the ballistic wave between the antenna and the target that is mixed with the backscattering response of the RC. A good agreement is obtained when compared with classical RCS measurement inside an anechoic chamber. This communication also highlights the potential stirrer positioning issues and their impact on the retrieved RCS accuracy.**

## I. Introduction

The Radar Cross Section (RCS) of an object is a crucial parameter in a wide range of applications including mainly military applications as stealth optimization and radar detection, but also civilian applications like RFID interrogation [1] and antenna characterization [2]. Its evaluation is usually performed within an Anechoic Chamber (AC) by measuring the reflection coefficient in either monostatic, quasi-monostatic or bistatic configurations.

In the last years, Reverberation Chambers (RCs) became a promising alternative testing facility for a wide range of electromagnetic applications including global parameter estimation sucha as  absorbing cross section [3] or antenna efficiency [4] and some more detailed feature analysis such as radiation pattern measurement [5]. Recently, the use of an RC as an alternative test environment to perform RCS measurement [6], [9]. This RCS characterization relies on the extraction of the ballistic wave backscattered by the target among the diffuse field backscattered by the RC itself. This approach presents some advantages by using a cheaper measurement setup compared to classical AC measurement. Also, and contrary to the method proposed in [7], no time gating is required, thus avoiding any Fourier Transform.

The method considered in this communication [6] implies a two-step process: first, the reflection coefficient of an antenna oriented towards the target is measured; second, the same measurement is performed without the target. Then, the difference between the two measurements leads to a quantity proportional to the object RCS through a regression process. Although not rotating during the measurement performed in [6], the stirrer is nonetheless present inside the RC. In this communication, we intend to show the impact of the stirrer when the two measurements (with and without target) are not performed for the same stirrer position. First, we introduce the theory of monostatic RCS measurement in an RC. In section III, the RCS of a canonical target (metallic plate) is estimated using this technique for a fixed stirrer position and compared with classical measurements in AC. Then, we show the impact of a shifted stirrer position on the accuracy of the RCS estimation. Finally, a conclusion ends this paper.

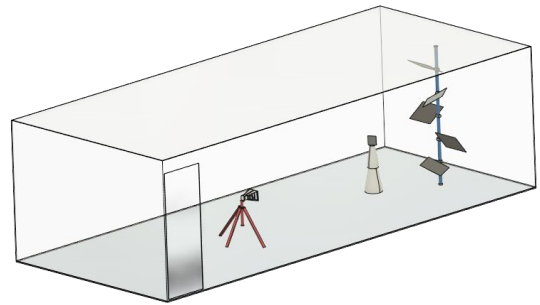

Fig. 1. Radar Cross Section measurement setup in Reverberation Chamber

## II. Measurement Method

### A. Measurement in the empty RC (without target)

Let us consider a single antenna located within the working volume of an RC and oriented towards the target (when added). Its reflection coefficient can be expressed as

$$S(f_0, \theta_{st}) = S_{FS}(f_0) + (1 - |S_{FS}(f_0)|^2)H(f_0, \theta_{st})\eta_{ant} \quad (1)$$

where $S_{FS}(f_0)$ is the free-space reflection coefficient of the antenna (that could be deduced as the mean of the reflection coefficient measured in RC over a stirrer rotation), and the second term is due to the backscattering response of the RC. It is composed of $H(f_0, \theta_{st})$ that represents the chamber transfer function at the antenna location and is supposed to follow a normal Gaussian distribution versus frequency. It is described at an arbitrary stirrer position $\theta_{st}$. This transfer function is weighted by $(1 - |S_{FS}(f_0)|^2)$ that represents the losses due to antenna mismatch, and $\eta_{ant}$ that stands for the antenna radiation efficiency. The stirrer position $\theta_{st}$ dependence will be our focus during this communication.

## B. Measurement in the loaded RC (with target)

Once the target is placed on the mast (that was already present for the empty measurement), the previous equation can be rewritten as

$$S^T(f_0, \theta_{st}^T) = S_{FS}(f_0) + C(f_0)\sqrt{\sigma^T(f_0)} + (1 - |S_{FS}(f_0)|^2)H^T(f_0, \theta_{st}^T)\eta_{ant} \quad (2)$$

The previous two terms in the right side of (1) are still present but the transfer function is changed because of the presence of the target and is now referred as $H^T(f_0, \theta_{st}^T)$ where the superscript $T$ stands for "target". The additional term $C(f_0)\sqrt{\sigma^T(f_0)}$ stands for the backscattering signal from the target towards the antenna. Especially, the complex quantity $C(f_0)$ describes the forward travelling wave from the antenna to the target and the backward travelling wave from the target to the antenna at the frequency $f_0$. The term $\sigma^T$ is the target RCS at the same frequency $f_0$. If the distance R between the antenna and the target is chosen such that it is greater than $2D^2/\lambda$ where D is the largest dimension of the target and $\lambda$ the minimum considered wavelength, the amplitude of $C(f_0)$ can be expressed thanks to the Friis equation:

$$|C(f_0)| = \frac{G_{ant}(f_0)\,\lambda_0}{(4\pi)^{3/2}R^2}(1 - |S_{FS}(f_0)|^2) \quad (3)$$

where $G_{ant}(f_0)$ is the antenna gain at the considered frequency. Assuming that the target size is much smaller than the distance R, the phase is supposedly constant along the target and this leads to the following simplified expression:

$$C(f_0) = |C(f_0)| \cdot exp\frac{-j2\pi f_0 2R}{c} \cdot exp(j\phi_0) \quad (4)$$

with $\phi_0$ an arbitrary constant phase. The phase variation corresponds to the wave propagation at the speed of light $c$ along the forward and backward paths of two times the distance $R$ between the measuring antenna and the target.

## C. Radar Cross Section expression

The expression that permits to extract the backscattering signal from the target which consequently leads to its $\sigma^T$, is found by computing the difference between the two previous measurements (with and without target):

$$S^T(f_0, \theta_{st}^T) - S(f_0, \theta_{st}) = (1 - |S_{FS}(f_0)|^2)(H^T(f_0, \theta_{st}^T) - H(f_0, \theta_{st}))\eta_{ant} + \sqrt{\sigma^T(f_0)}\frac{G_{ant}(f_0)\,\lambda_0}{(4\pi)^{3/2}R^2}(1 - |S_{FS}(f_0)|^2) \times exp\frac{-j2\pi f_0 2R}{c} \cdot exp(j\phi_0) \quad (5)$$

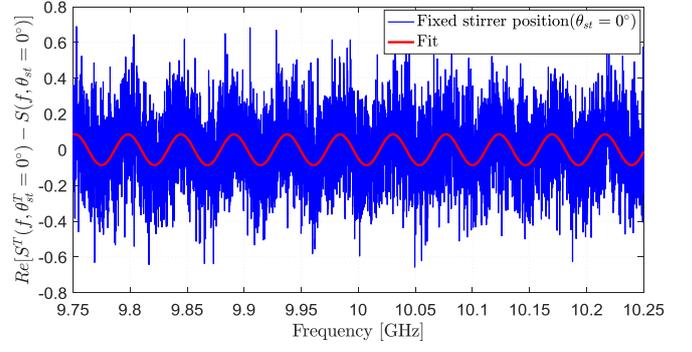

Fig. 2. Waveform versus frequency of the real part of the difference between the measured reflection coefficients of the loaded and empty cavities (for the same stirrer positon), at 0° target position.

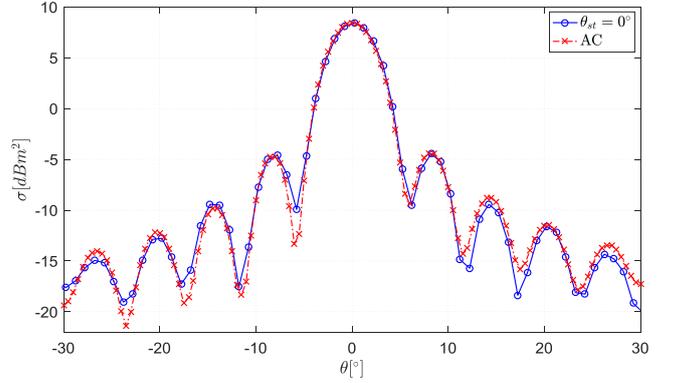

Fig. 3. RCS patterns at a frequency of 10 GHz, obtained from the difference between both measurements (with and without the target) for the same stirrer position ($\theta_{st}^T = \theta_{st} = 0°$), compared to the RCS mesured in AC.

The right-hand side of equation (5) consists in the addition of two terms. The first one is proportional to the difference between the two random variables, $H^T(f_0, \theta_{st}^T) - H(f_0, \theta_{st})$ following a centered Gaussian distribution (according to frequency or stirrer rotation). This assumption will permit to consider this term as an additive random noise and to extract the RCS from the second one. The latter contains the magnitude of the backscattered signal from the target to the illuminating antenna. Simplifying the equation (5), it becomes:

$$\frac{S^T(f_0, \theta_{st}^T) - S(f_0, \theta_{st})}{G_{ant}(f_0)\,\lambda_0(1 - |S_{FS}(f_0)|^2)} \propto \sqrt{\sigma^T(f_0)} \times exp\frac{-j2\pi f_0 2R}{c} \cdot exp(j\phi_0) \quad (6)$$

The amplitude estimation of the backscattered signal from the second term of the equation (6) allows deducing the RCS, all other parameters being determined.

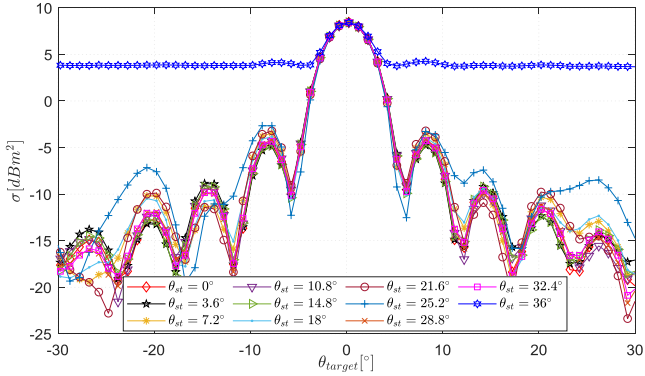

Fig. 4. RCS patterns at a frequency of 10 GHz, obtained from the difference between both measurements (with and without the target) for a stirrer position shift varying from 0° and 36°.

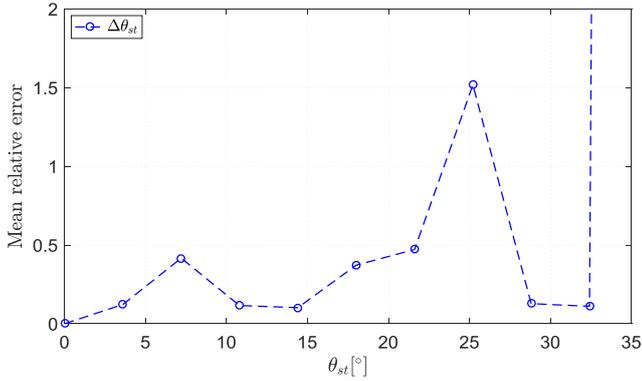

Fig. 5. Mean relative error between RCS patterns obtained with a stirrer position shift ($\theta_{st}^T \neq \theta_{st}$ with $\theta_{st}^T = 0°$). The mean of the relative error is calculated over the [-30°; 30°] target position range.

To begin our series of experiments about stirrer position effects, the difference between the two measurements is made for the same stirrer position, i.e. $\theta_{st}^T = \theta_{st}$, and the RCS pattern is extracted from the reflection coefficients difference. This approach is explicitly that of [6] and [8]. Secondly, the RCS pattern will be extracted from the difference between the reflection coefficients measured in RC loaded by the target, for the first stirrer position ($\theta_{st}^T = 0°$), and the reflection coefficient measured in the empty RC, for a shifted stirrer position ($\theta_{st}^T \neq \theta_{st}$). This set of experiments aims at observing the impact of changing stirrer position between both steps of the measurement process.

### III. METHOD VALIDATION ON A METALLIC PLATE

#### A. Measurement setup

A horn antenna, placed within the working volume of the RC, is oriented towards the target consisting of a metallic plate standing on the mast at a distance R = 2.95 m from the antenna (Fig. 1). The horn antenna is connected to a vector network analyzer (VNA) in order to measure its reflection coefficient. The horn antenna, the target and the stirrer are aligned, and these first two are centered at the same height.

Measurements are performed in 9.75 GHz – 10.25 GHz frequency range, therefore the distance R is large enough so that the ballistic wave can be considered locally as a plane wave at the target position. The considered frequency range is very high compared to the RC Lowest Useable Frequency (around 200 MHz), so that the field within the cavity can be considered as isotropic and uniform. The RCS of the metallic plate is evaluated over the azimuthal angle range θ = −30° to θ = 30° with an angular step of 1 degree. The horn-antenna is vertically polarized with an electric field along the height of the RC. Hence, the scattering parameter $S(f_0)$ is measured by the VNA (once calibrated at the antenna's connector level).

#### B. Extraction of the backscattered signal from the target

Fig. 1 shows the real part of the difference between the scattering parameters with and without target when the latter is orthogonal to the direct path from the antenna (this target position is mentioned as the 0° one), and the stirrer is at its initial position (labelled as 0° stirrer position). This signal shows indeed an oscillatory pattern corresponding to the second term of the right-hand side of (5). The signal of interest is then extracted from the noise in order to retrieve the RCS of the target [6]. The resulting fitted pattern appears as a red curve in Fig. 2.

#### C. Radar Cross Section extraction for a fixed stirrer position

The RCS pattern of a metallic plate, according to its azimuthal rotation, was estimated for a fixed stirrer position ($\theta_{st}^T = \theta_{st} = 0°$) and compared to the one obtained in a large anechoic chamber which provides a reference like measurement (Fig. 3). The obtained results have been normalized so that the maximal RCS value (corresponding to the target position 0°) is equal to the theoretical one $\sigma = \frac{4\pi S^2 f^2}{c^2}$ (with $S$ the target section and $c$ the light velocity).
A good agreement is obtained between both RCS patterns at a noticeable exception for the lowest levels of RCS. It permits us to conclude that the RCS pattern evaluated in a RC from the difference between two measurements (with and without the target) for the same stirrer position ($\theta_{st}^T = \theta_{st}$) leads to high accuracy results.

### IV. IMPACT OF A STIRRER POSITION SHIFT ON THE RCS ESTIMATION ACCURACY.

This section focuses on the impact of the mechanical stirrer position. In fact, the RCS pattern accuracy strongly depends on the choice of the stirrer position when computing the difference between the two measurements (with and without target). To show the impact of the shifted stirrer position in the evaluation of RCS pattern, we compute the difference between the $S_{11}$ parameter measured in RC loaded by the target at the initial stirrer position ($\theta_{st}^T = 0°$), and the $S_{11}$ measured in the empty RC at a different stirrer position ($\theta_{st}^T \neq \theta_{st}$).

## A. Radar Cross Section pattern for a shifted stirrer position

Fig. 4 shows the RCS patterns deduced from the difference between the two measurements for shifted stirrer positions. The curves related to stirrer positions near the first one (considered as a reference) remain similar. Higher differences are obtained for a shift of 25.2° and, for stirrer position shifts higher than 32.4°, it becomes impossible to retrieve RCS patterns.

## B. Estimation of the error according to the stirrer position

Fig. 5 shows the average relative error between the RCS patterns obtained for the same stirrer position at both measurement steps ($\theta_{st}^T = \theta_{st} = 0°$) and the ones related to a shifted stirrer position ($\theta_{st}^T \neq \theta_{st}$). The error remains low at low stirrer positions (below 20°) whereas the results become less reliable at higher shifts and the extremely high error from a shift of 36° reveals the impossibility to extract the RCS.

## V. CONCLUSION

This communication presents the RCS measurement of a metallic plate within an RC. The method is first compared to the traditional AC method and the results present a good agreement. Then, we highlight the impact of a stirrer position shift on the RCS estimation accuracy, i.e., when the stirrer position is not the same between the two needed measurements (with and without the target). The results show that the RCS pattern accuracy strongly decreases when the two considered stirrer positions are very different. However, for relatively close stirrer positions, the difference stays quite low. This behavior obviously depends on the stirrer itself, i.e., its geometry. Although the RC can be proposed as a valid and economically convenient alternative to the well-established AC for RCS measurements of canonical targets, we suggest some carefulness and the same stirrer position should be chosen for the two measurements, in order to increase the RCS pattern estimation accuracy.

ACKNOWLEDGMENT

This work was financially supported by the French Ministry of Defense (DGA/MRIS). This project was also supported in part by the European Union through the European Regional Development Fund, in part by the Ministry of Higher Education and Research, in part by the Région Bretagne, and in part by the Département d'Ille et Vilaine and Rennes Métropole, through the CPER Project SOPHIE/STIC & Ondes.